\documentclass[aps,prb,preprint,groupaddress]{revtex4}
\usepackage{txfonts}
\usepackage{braket}
\usepackage{bm}
\usepackage[pdftex]{graphicx}

\begin{document}
\title{
High-field Studies on Layered Magnetic and Polar Dirac Metals:\\Novel Quantum Transport Phenomena Coupled with Spin-valley Degrees of Freedom
}%

\author{Hideaki Sakai}\email{sakai@phys.sci.osaka-u.ac.jp}
\affiliation{Department of Physics, Osaka University, Toyonaka, Osaka 560-0043, Japan} 

\begin{abstract}
Recently, the interplay between the Dirac/Weyl fermion and various bulk properties, such as magnetism, has attracted considerable attention, since unconventional transport and optical phenomena were discovered.
However, the design principles for such materials have not been established well.
Here, we propose that the layered material $A$Mn$X_2$ ($A$: alkaline and rare-earth ions, $X$: Sb, Bi) is a promising platform for systematically exploring strongly correlated Dirac metals, which consists of the alternative stack of the $X^-$ square net layer hosting a 2D Dirac fermion and the $A^{2+}$-Mn$^{2+}$-$X^{3-}$ magnetic block layer.
In this article, we shall review recent high-field studies on this series of materials to demonstrate that various types of Dirac fermions are realized by designing the block layer.
First, we give an overview of the Dirac fermion coupled with the magnetic order in EuMnBi$_2$ ($A$=Eu).
This material exhibits large magnetoresistance by the field-induced change in the magnetic order of Eu layers, which is associated with the strong exchange interaction between the Dirac fermion and the local Eu moment.
Second, we review the Dirac fermion coupled with the lattice polarization in BaMn$X_2$ ($A$=Ba).
There, spin-valley coupling manifests itself owing to the Zeeman-type spin--orbit interaction, which is experimentally evidenced by the bulk quantum Hall effect observed at high fields.
\end{abstract}


\maketitle
%
\section{Introduction}
%
The Dirac/Weyl fermion in solids is a quasi-particle state described by a relativistic Dirac/Weyl equation.
Bismuth has been long known as a typical three-dimensional Dirac system\cite{Wolff1964, Fukuyma1970JPSJ, Fuseya2015JPSJ}, whereas graphene\cite{Castro2009RMP} and surface states of topological insulators\cite{Ando2013JPSJ, Hasan2010RMP} have been recently extensively investigated as two-dimensional ones.
Their unique physical properties, such as ultrahigh mobility beyond conventional semiconductors, have attracted much attention in terms of potential application as well as fundamental physics\cite{Vafek2014AnnuRev}.
In particular, reflecting the topologically nontrivial band structure, the relativistic quantum Hall effect (QHE) manifests itself in graphene\cite{Novoselov2005Nature, Zhang2005Nature} and topological insulator films\cite{Yoshimi2015NatCom}.
There, the half-integer quantization of the Hall plateaus and the zero-energy Landau level at the charge neutral Dirac point were experimentally discovered\cite{Novoselov2005Nature, Zhang2005Nature}, which are associated with the Berry phase of the Dirac fermion and hence have no analog in conventional 2D electron gas.
%
\par
%
More recently, the interplay between relativistic quasi-particles and various physical properties, such as magnetism, polarity (ferroelectricity), and electron correlation, has been of particular interest.
For instance, in magnetic topological insulator thin films, quantum Hall phenomena coupled with magnetic order were experimentally observed.
Typical examples are the quantum anomalous Hall effect showing a Hall plateau at zero field\cite{Chang2013Science, Checkelsky2014NatPhys} and an axion insulating state showing a zero Hall plateau\cite{Mogi2017NatMater}.
In addition to 2D thin films, 3D bulk materials, so called Dirac/Weyl semimetals\cite{Armitage2018RMP, Yan2017AnnuRev}, have also attracted significant interest for the coupling with a wide range of quantum phenomena.
In Weyl semimetals, the peculiar magnetic order and (polar) lattice structure give rise to a Weyl point (i.e., nondegenerate linearly crossing bands), leading to many unprecedented bulk properties, such as large anomalous Hall/Nernst effects\cite{Nakatsuji2015Nature, Nayak2016SciAdv, Suzuki2016NatPhys, Ikhlas2018NatPhys, Sumida2020ComMater, Sakai2018NatPhys}, magnetoresistance effects (chiral anomalies)\cite{Huang2015PRX, Xiong2015Science, Hirschberger2016NatMat, Kuroda2017NatMat}, and magnetooptical responses\cite{Higo2018NatPhoto, Okamura2020NatCom}.
In perovskite iridates, the highest mobility among the oxides was reported, which likely originates from the Dirac point formed in the vicinity of the Mott insulating (antiferromagnetic) state\cite{Fujioka2019NatCom}.
%
\par
\begin{figure}[tb]
\begin{center}
\includegraphics[width=.9\linewidth]{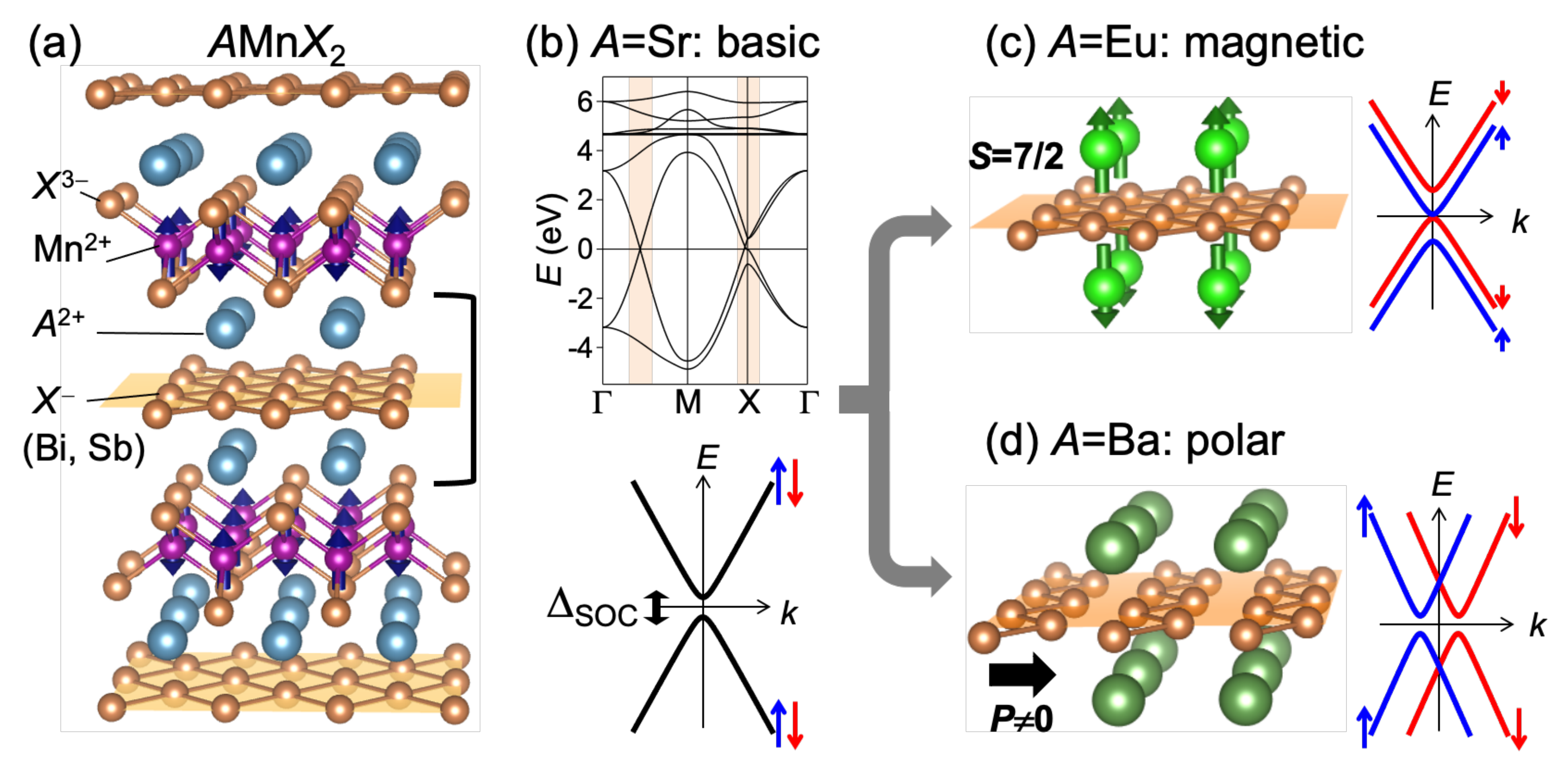}
\end{center}
\caption{(Color online)
(a) Lattice structure of $A$Mn$X_2$ ($A$: alkaline and rare-earth ions, $X$: Sb, Bi).
(b) Band structure for $A$=Sr, a basic material of $A$Mn$X_2$.
Upper panel: 2D band structure calculated by a tight-binding method for the $X^-$ square net coordinated with Sr (without the SOC).
Lower panel: schematic Dirac band on the $\Gamma$-M line obtained by a full first-principles calculation.
The energy gap is opened at the Dirac point by the SOC while keeping the spin-degenerate state.
(c) Square-net $X^-$ layer coordinated with magnetic Eu ions ($A$=Eu), where the Dirac band exhibits spin splitting due to the exchange interaction with local Eu$^{2+}$ spins ($S\!=\! 7/2$).
(d) Distorted (polar) $X^-$ layer coordinated with Ba ions ($A$=Ba), where the Dirac band exhibits valley-contrasting spin splitting due to the SOC with broken inversion symmetry (nonzero electric polarization $P$).
Reproduced with permission from Ref. \onlinecite{Sakai2021Butsuri} (\copyright 2021 Physical Society of Japan) and edited by the author.
}
\label{fig:intro}
\end{figure}

%
It is thus important to systematically explore the Dirac/Weyl semimetals to expand their variety as well as establish the underlying physics, which has never been demonstrated thus far.
This is because each Dirac/Weyl semimetal has its own peculiar magnetic and/or lattice structure and hence each material was discovered or predicted independently.
{\it How can we explore the Dirac/Weyl semimetals systematically?}
To address this issue, the layer structure consisting of the insulating block and conducting Dirac/Weyl fermion layers should be promising, since the Dirac/Weyl fermion layer can be controlled and modified by the coupling with the physical properties of the block layer, such as magnetic order, lattice distortion, and charge imbalance. 
This concept of material design (known as the block-layer concept or nanoblock integration) was first proposed for exploring high-$T_c$ cuprate superconductors\cite{Tokura1990JJAP}, followed by the application to a variety of strongly correlated systems (e.g., high-performance thermoelectric materials\cite{Koumoto2006MRS} and colossal magnetoresistive materials\cite{Kimura2000Annual}).
Here we propose that the layered material $A$Mn$X_2$ ($A$: alkaline and rare-earth ions, $X$: Sb, Bi) can realize the block-layer concept for the Dirac/Weyl semimetals, since $A$Mn$X_2$ consists of the alternative stack of the $X^-$ square net layer hosting quasi 2D Dirac fermions and the Mott insulating $A^{2+}$-Mn$^{2+}$-$X^{3-}$ block layer [Fig. \ref{fig:intro}(a)].
%
\par
%
In this review, we give an overview of the novel physics of Dirac fermions coupled with the magnetic order and lattice polarization in this series of compounds.
Owing to the high-mobility features of the Dirac fermion layer spatially separated from the block layer, high-field measurements using a pulsed magnet enable us to observe quantum oscillations (and even quantum Hall effects) with high accuracy.
This plays a vital role in revealing the electronic structure as well as transport properties of the correlated Dirac fermion in $A$Mn$X_2$.
%
%
\section{Measurements at High Fields}
%
We performed transport and magnetization measurements up to $\sim$55 T using a non-destructive mid-pulse magnet equipped with a 900 kJ capacitor bank at the International MegaGauss Science Laboratory at the Institute for Solid State Physics, The University of Tokyo.
All the measurements were performed on single crystals of $A$Mn$X_2$ grown by a high-temperature solution growth technique\cite{Canfield2001JCG}.
The measurement temperature range was 1.4--150 K.
The resistivity was measured by a lock-in technique at 100 kHz with a typical AC excitation of 1--10 mA, where the field direction was controllable by using a sample probe equipped with a rotating stage.
The tilt angle of the field was determined by two pick-up coils fixed parallel and perpendicular to the sample stage.
As detailed in Sects. \ref{sec:EMB_Rzz} and \ref{sec:BMX_QHE}, the measurement of the angular dependence of quantum oscillation is essential for unveiling the microscopic properties of the Dirac fermions.
The magnetization at high fields was measured by the induction method, using coaxial pickup coils.
%
\section{Basic Electronic and Lattice Structures of $A$Mn$X_2$}
\label{sec:AMnX2}
%
The $X^-$ layer in $A$Mn$X_2$ works as a conducting Dirac fermion layer and dominates the transport properties [shaded layer in Fig. \ref{fig:intro}(a)].
From the tight-binding calculation, the band structure for the $X^-$ square net coordinated with the $A$ site ions exhibits the spin-degenerate 2D linear bands, which cross on the $\Gamma$-M line and around the X point [upper panel of Fig. \ref{fig:intro}(b)].
On the other hand, the $A^{2+}$-Mn$^{2+}$-$X^{3-}$ layer works as a block layer, where the Mn layer is a robust Mott insulator with the antiferromagnetic (AFM) order around room temperature irrespective of the $A$ and $X$ species.
The first-principles calculation for $A$Mn$X_2$ indeed predicted that quasi-2D massive Dirac bands are formed as a bulk band\cite{Lee2013PRB,Farhan2014JPC}, where the energy gap at the charge neutral Dirac point originates from the spin-orbit coupling (SOC) of the $X^-$ ion [lower panel of Fig. \ref{fig:intro}(b)].
Among the $A$Mn$X_2$ materials, SrMnBi$_2$ ($A$=Sr, $X$=Bi) first drew attention in 2011\cite{Park2011PRL,Wang2011PRB,WangPetrovic2011PRB}.
This material is regarded as a basic material of $A$Mn$X_2$, since it exhibits simple spin-degenerate quasi-2D Dirac bands, as clearly observed along the $\Gamma$-M line by angle-resolved photoemission spectroscopy (ARPES)\cite{Park2011PRL,Feng2014SR, Ishida2016PRB}.
%
\par
%
In SrMnBi$_2$, the AFM order of the Mn layer apparently has little influence on the Dirac fermion\cite{Nishiyama2021PRB}; no resistive anomalies were observed at the magnetic transition temperature\cite{Park2011PRL,Wang2011PRB,WangPetrovic2011PRB}.
In order to enhance the impact of the block layer on the Dirac fermion, the substitution of the $A$ sites would be most effective, since the $A$ site located adjacent to the $X^-$ layer can affect its electronic and lattice structures directly. 
In Sect. \ref{sec:EMB}, we first review the Dirac fermion coupled with the magnetic order in EuMnBi$_2$ [Fig. \ref{fig:intro}(c)], where nonmagnetic Sr in the $A$ site is substituted with magnetic Eu hosting local spin $S\!=\! 7/2$.
High-field transport measurements have revealed that the AFM order of Eu layers strongly affects not only the quantum transport but also the Dirac band via the exchange interaction.
Next, in Sect. \ref{sec:BMX}, we review the Dirac fermion coupled with lattice polarization $P$ in BaMn$X_2$ [Fig. \ref{fig:intro}(d)], where Sr is substituted with Ba hosting the larger ionic radius to induce polar lattice distortion in the $X^-$ layer.
The SOC together with the broken inversion symmetry gives rise to the spin polarization dependent on Dirac valleys, which is clearly reflected in the bulk half-integer QHE observed at high fields.
In Sect. \ref{sec:summary}, we will give a summary of this article.
%
\par
%
Note here that $A$Mn$X_2$ has another crystal structure with the staggered arrangement of $A$ sites below and above the $X^-$ layer, as is the case for CaMn$X_2$~\cite{Wang2012PRB, He2017PRB}, YbMn$X_2$~\cite{Borisenko2019NatCom, Wang2018PRM, Kealhofer2018PRB}, $A$MnSb$_2$ ($A$=Sr, Eu)\cite{Liu2017NatMater, Ramankutty2018SciPost, You2018CAP, Yi2017PRB, Soh2019PRB}, and so on.
For this layer structure, the nonsymmorphic symmetry inherent in the space group $P4/nmm$ or $Pnma$ results in not a simple 2D Dirac band, but complex bands hosting a large nodal line\cite{Hyun2018PRB}, which is beyond the scope of this review.
%
\section{Dirac Fermion Coupled with Magnetic Order in EuMnBi$_2$}
\label{sec:EMB}
%
\subsection{Large magnetoresistance effect of high-mobility Dirac fermions}
%
\begin{figure}[tb]
\begin{center}
\includegraphics[width=.8\linewidth]{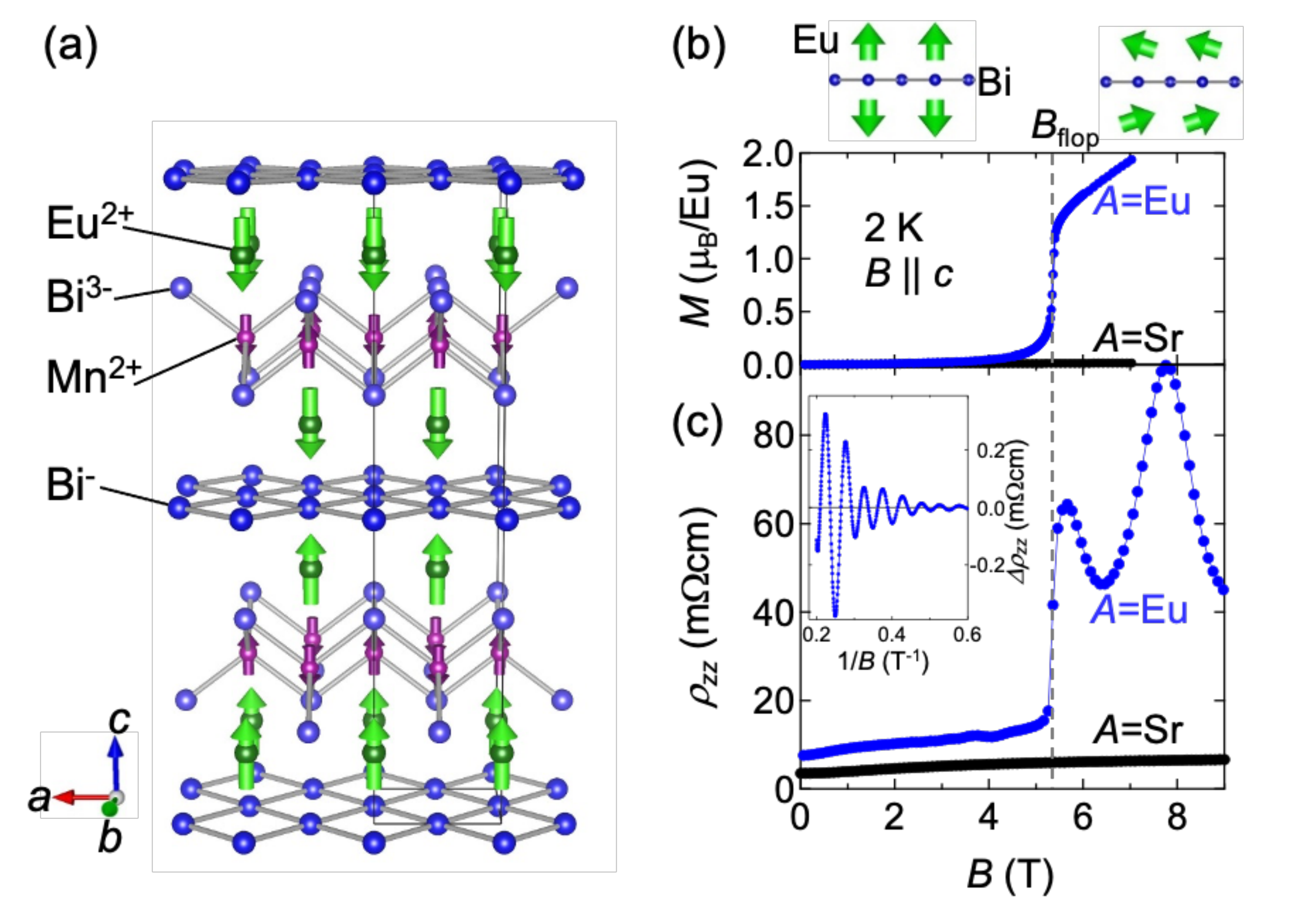}
\end{center}
\caption{(Color online)
(a) Lattice and magnetic structures for EuMnBi$_2$ at zero field\cite{Masuda2020PRB,Zhu2020PRR}, together with the formal valence of each ion.
Magnetic field $B$ dependences of (b) magnetization $M$ and (c) interlayer resistivity $\rho_{zz}$ for EuMnBi$_2$ ($A$=Eu) and SrMnBi$_2$ ($A$=Sr) at 2 K for the field parallel to the $c$-axis.
The directions of Eu$^{2+}$ moments at $B\!<\!B_{\rm flop}$ and $B_{\rm flop}\!<\!B$ are shown in upper panels of (b).
The inset of (c) shows the oscillating component of $\rho_{zz}$ ($\Delta \rho_{zz}$) versus $1/B$ in the low-field AFM phase (below $B_{\rm flop}$) for EuMnBi$_2$.
Reproduced with permission from Refs. \onlinecite{Sakai2021Butsuri} (\copyright 2021 Physical Society of Japan) and \onlinecite{Masuda2016SA}  (\copyright 2016 The authors) and edited by the author.
}
\label{fig:EMB_MR}
\end{figure}
%
For EuMnBi$_2$, Eu layers exhibit the AFM order below $T_N$=22 K at 0 T\cite{May2014PRB, Masuda2016SA}, where Eu moments order ferromagnetically within the $ab$ plane and align along the $c$-axis in the sequence of up-up-down-down [Fig. \ref{fig:EMB_MR}(a)]\cite{Masuda2016SA, Masuda2020PRB, Zhu2020PRR}.
The Bi square net intervenes between Eu layers with magnetic moments up and down, reminiscent of a natural spin-valve-like structure.
Importantly, the AFM states of Eu layers can be controlled by manipulating the external magnetic field along the $c$-axis.
The magnetization at 2 K shows a clear metamagnetic (spin-flop) transition at $B\!=\! B_{\rm flop}$ ($\sim\!5.3$ T), corresponding to the reorientation of Eu moments to be perpendicular to the field [Fig. \ref{fig:EMB_MR}(b)].
The interlayer resistivity $\rho_{zz}$ shows a large jump at $B_{\rm flop}$; $\rho_{zz}$ above $B_{\rm flop}$ is 5-10 times as high as that at 0 T [Fig. \ref{fig:EMB_MR}(c)].
The marked oscillating component above $B_{\rm flop}$ arises from the Shubnikov--de Haas (SdH) oscillation [see Fig. \ref{fig:EMB_highB}(c) for details].
Such a field dependence of $\rho_{zz}$ is totally different from that for SrMnBi$_2$, which shows a minimal magnetoresistance effect up to 9 T.
In EuMnBi$_2$, the clear SdH oscillation was observed not only above $B_{\rm flop}$ but also below $B_{\rm flop}$, which is discernible above as small as 1.5 T [inset of Fig. \ref{fig:EMB_MR}(c)].
This indicates that the transport of the Dirac fermion is largely tunable by changing the AFM order of Eu layers, while keeping the high mobility ($\sim$15,000 cm$^2$/Vs at 2 K)\cite{Masuda2016SA, Tsuruda2021AdvFunctMater}.
%
\subsection{Quantum transport coupled with the AFM order of Eu layers}
\begin{figure}[tb]
\begin{center}
\includegraphics[width=.9\linewidth]{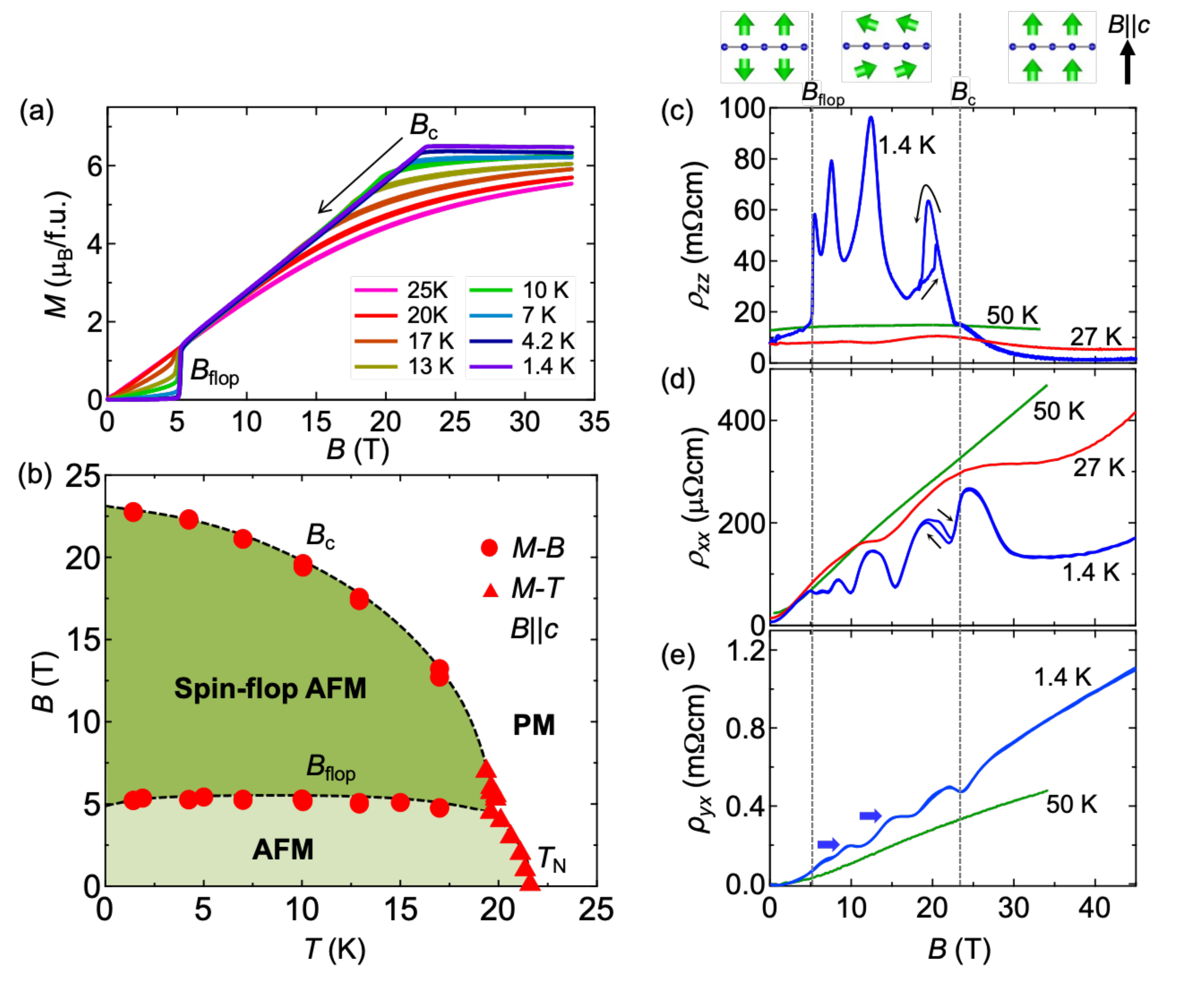}
\end{center}
\caption{(Color online)
(a) Field ($B$) dependence of magnetization ($M$) for EuMnBi$_2$ up to 33 T at selected temperatures ($B||c$), measured with a pulsed magnet.
(b) Magnetic phase diagram for the Eu sublattice as functions of $B$ and temperature ($T$).
PM and AFM denote the paramagnetic and low-field antiferromagnetic phases, respectively.
$B_{\rm flop}$ and $B_{\rm c}$ denote the transition fields to the spin-flop AFM and PM (forced ferromagnetic) phases, respectively.
$B$ dependences of (c) $\rho_{zz}$,  (d) $\rho_{xx}$, and (e) Hall resistivity $\rho_{yx}$ for EuMnBi$_2$ up to $\sim$45 T at selected temperatures  ($B||c$).
The directions of the Eu$^{2+}$ moments adjacent to the Bi layer are schematically presented on the top of panel (c).
Horizontal arrows in panel (e) indicate plateau-like structures of $\rho_{yx}$ at 1.4 K.
Reproduced with permission from Ref. \onlinecite{Masuda2016SA}  (\copyright 2016 The authors) and edited by the author.
}
\label{fig:EMB_highB}
\end{figure}

%
The quantum transport phenomena of the high-mobility Dirac fermion are clarified by high-field measurements up to 55 T.
We first show the overall field and temperature dependences of the AFM order of Eu layers.
As shown in Fig. \ref{fig:EMB_highB}(a), the magnetization at 1.4 K increases linearly with the field above $B_{\rm flop}$, followed by saturation above $B_{\rm c}\!\sim\! 22$ T.
The saturated moment is close to 7$\mu_{\rm B}$/Eu, reflecting localized Eu 4$f$ electrons.
The temperature dependences of $B_{\rm flop}$ and $B_{\rm c}$ deduced from Fig. \ref{fig:EMB_highB}(a) are plotted in Fig. \ref{fig:EMB_highB}(b), forming a typical phase diagram for an anisotropic antiferromagnet for the field along the magnetization-easy axis.
%
\par
%
Figures \ref{fig:EMB_highB}(c)--(e) show the field dependence of the transport properties for EuMnBi$_2$ at selected temperatures.
As expected from the low-field data in Fig. \ref{fig:EMB_MR}(c), $\rho_{zz}$ is markedly dependent on the AFM states of Eu layers [Fig. \ref{fig:EMB_highB}(c)].
Above $T_{\rm N}$ (at 27 and 50 K), $\rho_{zz}$ is almost independent of the field, except for SdH oscillations at 27 K.
Below $T_{\rm N}$ (at 1.4 K), $\rho_{zz}$ significantly increases above $B_{\rm flop}$, where the giant SdH oscillation reaching $\Delta \rho_{\rm osc}/\rho\!\sim$50\% was observed.
This high-$\rho_{zz}$ state is terminated at $B_{\rm c}$, above which $\rho_{zz}$ is substantially reduced.
The in-plane resistivity $\rho_{xx}(B)$ exhibits a large positive (transverse) magnetoresistance effect, irrespective of the Eu magnetic order [Fig. \ref{fig:EMB_highB}(d)].
At 50 K, the $\rho_{xx}(B)$ profile is strikingly $B$-linear without saturation up to 35 T, resulting in the magnetoresistance ratio of $\rho(B\!=\!35\ {\rm T})/\rho(0)\!\sim$ 2,000\%.
Such a large $B$-linear magnetoresistance is occasionally observed in Dirac semimetals\cite{Liang2014NatMat, He2014PRL, Narayanan2015PRL, Kushwaha2015APLMat, Novak2015PRB}.
At lower temperatures, the SdH oscillation is superimposed; at 1.4 K, the magnitude of SdH oscillation is particularly enhanced in the spin-flop AFM phase ($B_{\rm flop}\!<\!B\!<\!B_{\rm c}$), similarly to $\rho_{zz}$.
The corresponding SdH oscillation in the Hall resistivity $\rho_{yx}$ results in plateau-like structures at 1.4 K, as denoted by horizontal arrows in Fig. \ref{fig:EMB_highB}(e).
We shall later analyze $\rho_{yx}$ plateaus in detail in terms of the bulk QHE occurring in stacked Dirac fermion (Bi$^-$) layers (See Sect. \ref{sec:BMX_QHE}).
%
\par
%
%
As an interesting transport phenomenon at high fields, we note that the peak structure in $\rho_{zz}$ and $\rho_{xx}$ around 20 T shows a sizable hysteresis between the field-increasing and field-decreasing runs [Figs. \ref{fig:EMB_highB}(c) and \ref{fig:EMB_highB}(d)].
Since no clear anomaly is discerned in the magnetization curve around 20 T [Fig. \ref{fig:EMB_highB}(a)], Eu moments play a minor role.
Instead, a possible electronic phase transition of Dirac fermions at high fields is implied, although the precise origin remains unclear at present.
%
\subsection{Exchange coupling between Dirac fermion and Eu spin: AFM-order-dependent $g$ factor}
\label{sec:EMB_Rzz}
%
\begin{figure}[tb]
\begin{center}
\includegraphics[width=.4\linewidth]{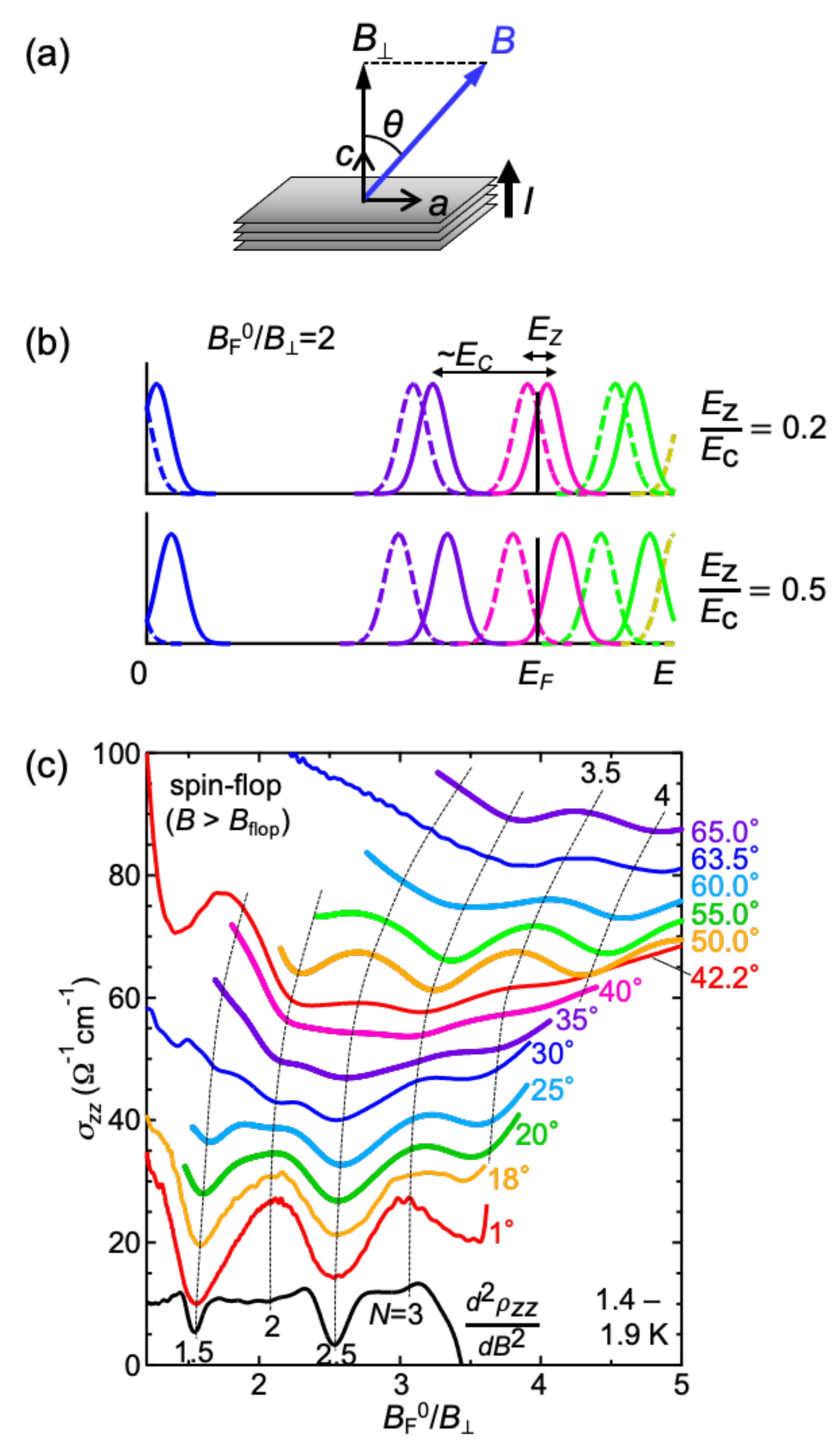}
\end{center}
\caption{(Color online)
(a) Geometry of the $\rho_{zz}$ measurement for the magnetic field rotated in the $a$-$c$ plane, where $\theta$ is an angle between the field and the $c$-axis.
(b) Schematic of the density of states of spin-split Landau levels for a 2D Dirac fermion as a function of energy $E$ at $B_F^0/B_\perp=2$, where $B_F^0$ denotes the SdH frequency for the field parallel to the $c$-axis.
$E_{\rm Z}/E_{\rm c}$, where $E_{\rm Z}$ is the Zeeman energy and $E_{\rm c}$ the cyclotron energy, can be tuned by tilting the field; $E_{\rm Z}/E_{\rm c}\!=\! 0.2$ (top) and 0.5 (bottom).
(c) $\sigma_{zz}$ versus $B_F^0/B_\perp$ at $\theta$=1$^{\circ}$-~65$^{\circ}$ in the spin-flop AFM phase ($B_{\rm flop}\!<\!B\!<\!B_{\rm c}$).
The curves at $\theta\ge 18^{\circ}$ are shifted upward for clarity.
At the bottom of the panel, the second field derivative $d^2\rho_{zz}/dB^2$ at $\theta\!=\! 1^\circ$ is shown.
Vertical dotted lines are guides to the eye showing the positions of the maxima and minima of the SdH oscillation, where $N$ denotes the Landau index.
Reproduced with permission from Ref. \onlinecite{Masuda2018PRB} (\copyright 2018 American Physical Society) and edited by the author.
}
\label{fig:EMB_Rzz}
\end{figure}
%
Having clarified that the AFM order has a marked impact on transport properties, we now discuss how and to what extent the Dirac band dispersion is affected.
In this light, Landau level quantization provides much information, since it exhibits energy splitting not only due to electron--electron interactions, but also due to Zeeman and exchange interactions.
The analysis on the splitting enables the determination of the band parameters of 2D systems, as demonstrated for conventional semiconductor heterostructures\cite{Ando1982RMP}, and more recently, for graphene\cite{Zhang2006PRL,Young2012NatPhys} and several Dirac semimetals\cite{Jeon2014NatMater,Xiang2015PRL,Liu2016NatCom}.
In this subsection, we overview the features of Landau level splitting for EuMnBi$_2$, which was revealed by systematic high-field measurements under tilted magnetic fields\cite{Masuda2018PRB}.
Magnetic field rotation is important for 2D systems, since the ratio of the cyclotron energy $E_{\rm c}$ to the Zeeman energy $E_{\rm Z}$ can be tuned by changing the tilt angle of the field from the normal to the 2D plane ($\theta$).
Note here that $E_{\rm c}$ is proportional to $B_\perp=B\cos\theta$ [the field component perpendicular to the 2D plane, see Fig. \ref{fig:EMB_Rzz}(a)], whereas $E_{\rm Z}$ is proportional to $B$ (total field).
For clarifying the fine structures of split Landau levels, we adopted the measurements of interlayer resistivity ($\rho_{zz}$).
This is because the highly resistive $\rho_{zz}$ has a much higher S/N ratio than $\rho_{xx}$.
By combining these techniques, we have elucidated the variation of the Dirac bands upon the AFM order in EuMnBi$_2$.
%
\par
%
Figure \ref{fig:EMB_Rzz}(c) displays the $\theta$ dependence of interlayer conductivity $\sigma_{zz}=1/\rho_{zz}$, which corresponds to the density of states of Landau levels, in the spin-flop AFM phase ($B_{\rm flop}\!<\!B\!<\!B_c$).
The horizontal axis of Fig. \ref{fig:EMB_Rzz}(c) is the normalized filling factor $B_F^0/B_\perp$\cite{Lukyanchuk2006PRL}, where $B_F^0$(=19.3 T) is the SdH frequency for the field along the $c$-axis ($\theta\!=\! 0^\circ$).
At $\theta$=1$^{\circ}$, $\sigma_{zz}$ shows the minima at around $B_F^0/B_\perp\!=\!1.5$, 2.5, and 3.5.
Since the $\sigma_{zz}$ minima correspond to the energy gap between the neighboring Landau levels\cite{Druis1998PRL,Kuraguchi2000PhysicaE,Kawamura2001PhysicaB}, they occur at $B_F^0/B_\perp\!=\! N + 1/2 - \gamma$, where $N$ is the Landau index and $\gamma$ is the phase factor expressed as $\gamma\!=\! 1/2 - \phi_B/2\pi$ with $\phi_B$ being the Berry's phase\cite{Mikitik1999PRL}.
Hence, the $\sigma_{zz}$ minima at half-integral $B_F^0/B_\perp$ indicate $\gamma\!\sim\!0$, i.e., the nontrivial $\pi$ Berry's phase in EuMnBi$_2$.
When $\theta$ increases, the frequency of the SdH oscillation increases proportionally to $1/\cos\theta$, consistent with the quasi-2D Fermi surface.
This results in the almost $\theta$-independent oscillation period when plotted as a function of $B_F^0/B_\perp$, as shown by the vertical dotted lines up to $\theta\sim 50^{\circ}$ in Fig. \ref{fig:EMB_Rzz}(c).
For $\theta\!\ge\! 55^{\circ}$, however, the frequency gradually deviates from the $1/\cos\theta$ scaling presumably owing to a weak warping of the cylindrical Fermi surface.
%
\par
%
What is more important is that the amplitude of SdH oscillation significantly varies with $\theta$.
With increasing $\theta$, the amplitude gradually decreases and reaches nearly zero at $\theta$=35$^\circ$-40$^\circ$.
Above $\theta\!=\! 40^\circ$, the amplitude again increases but with an inverted phase.
The observed $\theta$ dependence of the SdH oscillation is well reproduced by considering the spin splitting of Landau levels due to $E_{\rm Z}$ as follows\cite{Si100_spinsucceptibility_Fang1968, AlAs_spinsucceptibility_Vakili2004, ZnO_spinsucceptibility_Tsukazaki2008}.
When $E_{\rm Z}/E_{\rm c}$ is smaller than unity [e.g., $E_{\rm Z}/E_{\rm c}\!=\! 0.2$ in the upper panel of Fig. \ref{fig:EMB_Rzz}(b)], the Landau level exhibits a weak spin splitting, as barely discernible in the profile of $d^2\rho_{zz}/dB^2$ at $\theta\!\sim\! 1^\circ$ [Fig. \ref{fig:EMB_Rzz}(c)].
With increasing $E_{\rm Z}/E_{\rm c}$ by tilting the field, the magnitude of the spin splitting increases and the amplitude of SdH oscillation decreases.
Around $\theta\!=\! 40^\circ$, the amplitude reaches the minimum, corresponding to $E_{\rm Z}/E_{\rm c}\!=\! 0.5$ [lower panel of Fig. \ref{fig:EMB_Rzz}(b)].
A further increase in $E_{\rm Z}/E_{\rm c}$ leads to the crossing of the neighboring Landau levels with opposite spins.
This results in not only the amplitude enhancement but also the phase inversion of SdH oscillation, as observed at $\theta\!>\! 50^\circ$.
%
\par
%
\begin{figure}[tb]
\begin{center}
\includegraphics[width=\linewidth]{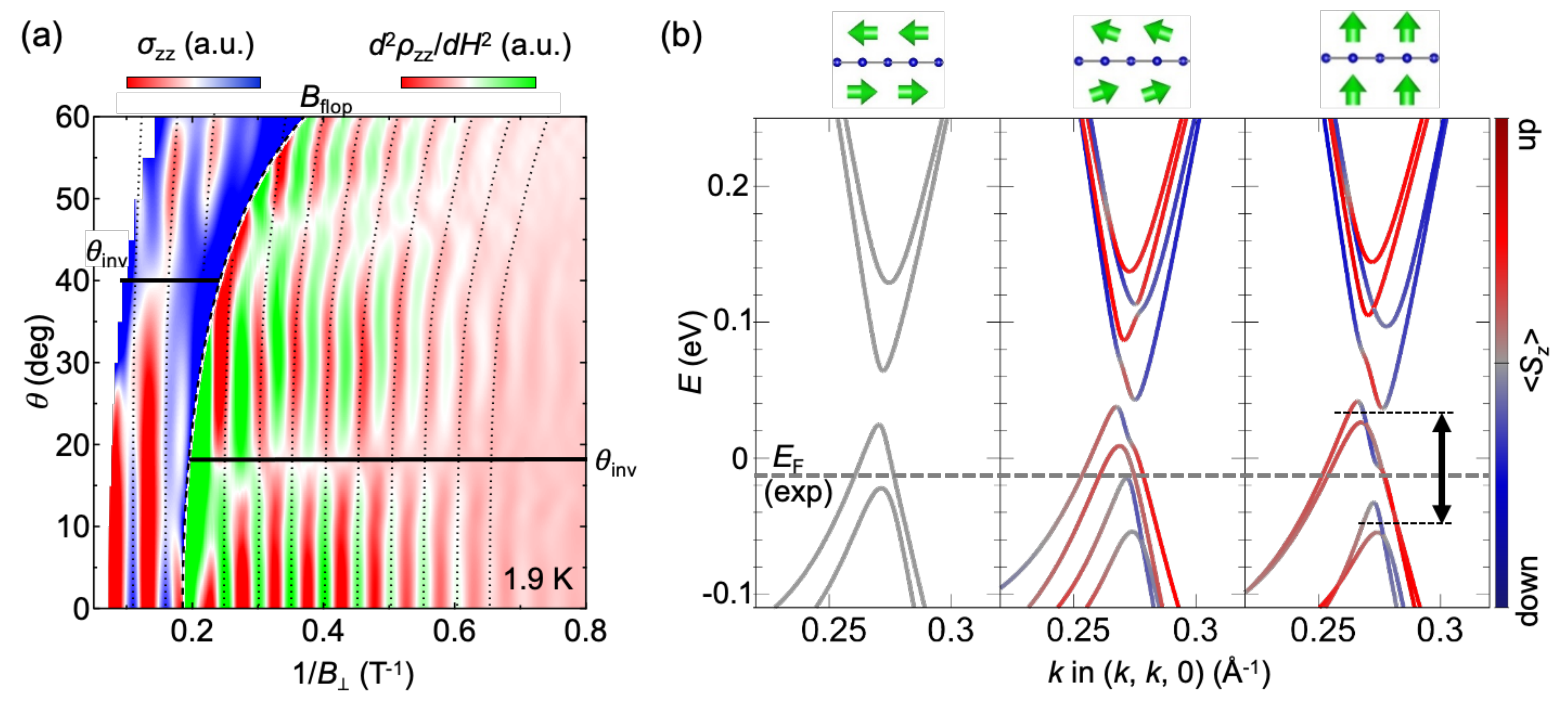}
\end{center}
\caption{(Color online)
(a) Color plot of $d^2\rho_{zz}/dB^2$ (low-field AFM phase; right) and $\sigma_{zz}$ (high-field spin-flop AFM phase; left) as functions of $1/B_\perp$ and $\theta$.
The two phases are separated by $B_{\rm flop}$ (vertical dashed curve).
The horizontal line denotes $\theta_{\rm inv}$ for each phase.
(b) Calculated Dirac bands along the $\Gamma$-M line for various magnetic states in EuMnBi$_2$.
Eu moment is oriented along the $a$-axis in the left panel, while it is inclined at an angle of $\sim$47$^\circ$ to the $c$-axis on the $ac$ plane in the middle panel.
The right panel corresponds to the forced ferromagnetic state.
The spin polarization $\langle S_z\rangle$ of each band is represented by red (up) and blue (down) colors.
The magnitude of spin-splitting energy for the forced ferromagnetic state is denoted by the vertical arrow in the right panel ($\sim$ 80 meV).
The Fermi energy $E_{\rm F}$ estimated from the experimental SdH oscillation is denoted by the horizontal dashed line.
Reproduced with permission from Ref. \onlinecite{Masuda2018PRB} (\copyright 2018 American Physical Society) and edited by the author.
}
\label{fig:EMB_color}
\end{figure}
%
To overview the $\theta$ dependence of the SdH oscillation, we show a contour plot of its amplitude as functions of $1/B_\perp$ and $\theta$ in Fig. \ref{fig:EMB_color}(a).
The plot is divided into two regions by $B_{\rm flop}$; the left side corresponds to the spin-flop AFM phase (above $B_{\rm flop}$), whereas the right side the AFM phase (below $B_{\rm flop}$).
There are several common features in the SdH oscillation for both phases; the period of the SdH oscillation is nearly independent of $\theta$ when plotted versus $1/B_\perp$, reflecting a quasi-2D Fermi surface.
Furthermore, owing to the spin splitting of Landau levels, the phase of the SdH oscillation is inverted around $\theta_{\rm inv}$ (horizontal line).
However, $\theta_{\rm inv}$ substantially differs between the two phases: $\theta_{\rm inv}\!\sim\! 18^\circ$ for the AFM phase, whereas $\theta_{\rm inv}\!\sim\! 40^\circ$ for the spin-flop AFM phase.
Since $\theta\!=\! \theta_{\rm inv}$ corresponds to $E_{\rm Z}/E_{\rm c}\!=\! 0.5$ [see the lower panel of Fig. \ref{fig:EMB_Rzz}(b)], we obtain $\cos\theta_{\rm inv}\!=\! g^\ast m_{\rm c}/m_0$ by substituting $E_{\rm Z}\!=\! g^\ast\mu_{\rm B}B$ and $E_{\rm c}\!=\! e\hbar B_\perp /m_{\rm c}$\cite{note_E_c}, where $g^\ast$ is the effective $g$ factor and $m_0$ is the bare electron mass.
Noting that the $m_{\rm c}/m_0$ value estimated from the temperature dependence of the SdH oscillation is almost independent of the type of AFM order\cite{Masuda2018PRB}, we can conclude that the difference in $\theta_{\rm inv}$ originates mostly from that in $g^\ast$ between the two AFM phases.
%
\par
%
First-principles calculations have revealed a marked dependence of the band structure on the magnetic state of the Eu layers.
Figure \ref{fig:EMB_color}(b) displays the Dirac bands near $E_{\rm F}$ calculated for various magnetic states.
For the AFM state with Eu moments lying perfectly within the plane (left panel), the Dirac bands are spin-degenerate (gray color)\cite{note_cell_doubling}.
However, when Eu moments are inclined toward the $c$-axis, the Dirac bands exhibit clear spin splitting (middle panel), where the red color corresponds to the spin up, whereas the blue color the spin down.
As the net magnetization (i.e., the canting of the Eu moment) increases, the spin splitting of the bands progressively evolves and reaches 50$-$80 meV for the forced ferromagnetic state (right panel).
This large spin splitting is caused by the exchange interaction between the Dirac fermion and the local Eu moment; $E_{\rm ex}\!=\! J\langle S\rangle \!=\! J\chi H/g_{J}$, where $J$ is the exchange integral, $\langle S\rangle$ is the component of the Eu$^{2+}$ spin along the field, $g_{J}(=2)$ is the Land\'e $g$ factor for Eu$^{2+}$, and $\chi$ is the magnetic susceptibility.
In the AFM phase, $\chi$ corresponds to a small parallel susceptibility and hence $E_{\rm ex}$ is negligibly small.
However, in the spin-flop AFM phase, the Eu spin axis changes to be transverse to the field and $\chi$ changes to a large transverse susceptibility [Figs. \ref{fig:EMB_MR}(b) and \ref{fig:EMB_highB}(a)].
As a result, $E_{\rm ex}$ is comparable to $E_{\rm Z}$ in the latter phase\cite{note_Eex_Ez}, where $g^\ast$ should be renormalized by the significant contribution of $E_{\rm ex}$ to the total Landau level splitting.
Thus, the observed apparent change in $g^\ast$ upon the AFM order is a firm evidence of the strong coupling between the Dirac band and the local Eu moment in EuMnBi$_2$.
More recently, it has been revealed that the exchange coupling strongly affects the thermoelectric phenomena (Seebeck and Nernst effects) as well\cite{Tsuruda2021AdvFunctMater}.
These facts indicate that various electrical and thermal transport phenomena are tunable by controlling the Dirac band (splitting) via the magnetic Eu layers in this material.
%
\section{Dirac Fermion Coupled with Electric Polarization in BaMn$X_2$}
\label{sec:BMX}
%
\subsection{Polar lattice structure}
\begin{figure}[tb]
\begin{center}
\includegraphics[width=\linewidth]{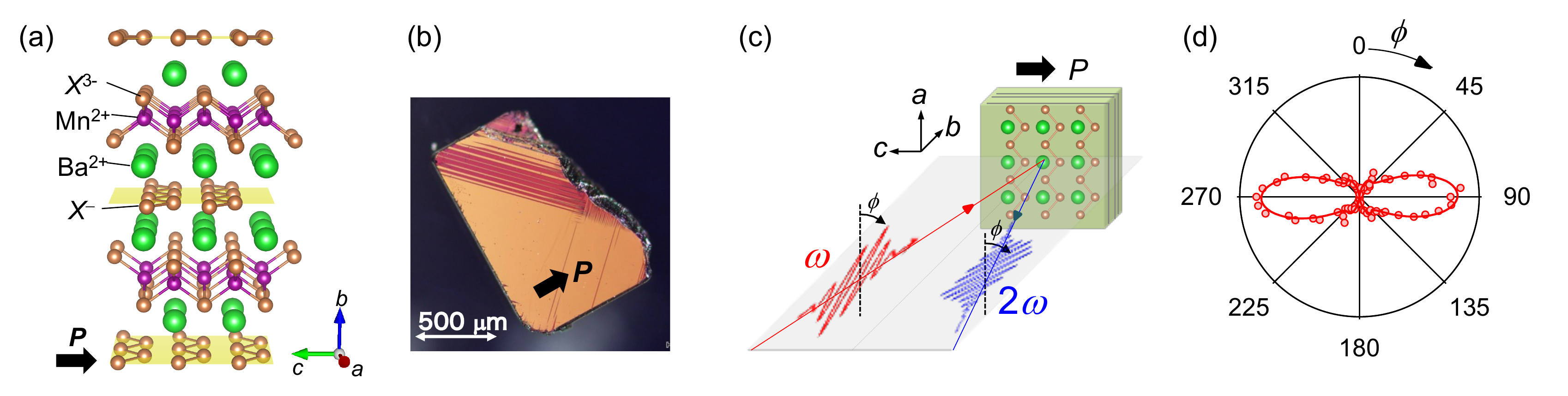}
\end{center}
\caption{(Color online)
(a) Crystal structure of BaMn$X_2$ ($X$=Sb, Bi), where $X^-$ 1D zigzag chains coordinated with Ba ions generate lattice polarization ($\bm{P}$) along the $c$-axis.
(b) Polarized microscopy image of an as-grown crystal surface ($X$=Sb).
(c) Schematic image of SHG measurement at nearly normal incidence ($\sim1^{\circ}$).
The angle $\phi$ denotes the polarization direction of the incident fundamental light ($\omega$) and the emitted SHG light ($2\omega$).
(d) Polarization analysis of the SHG signal ($X$=Bi).
The red solid curve is a fitted result assuming the $mm2$ symmetry.
Reproduced with permission partly from Refs. \onlinecite{Sakai2020PRB} (\copyright 2020 American Physical Society) and \onlinecite{Kondo2021CommunMater} (\copyright 2021 The authors) and edited by the author.
}
\label{fig:BMX_lattice}
\end{figure}

%
Figure \ref{fig:BMX_lattice}(a) schematically shows the lattice structure of BaMn$X_2$ ($X$=Sb, Bi), where the $X^-$ square net is slightly distorted to a zigzag chainlike structure.
Although the structure of BaMn$X_2$ was previously reported to be tetragonal\cite{Liu2015SR,Huang2016PNAS}, our recent single-crystal X-ray structural analysis has revealed that BaMn$X_2$ has a weak orthorhombic distortion.\cite{Sakai2020PRB, Kondo2021CommunMater}
This is also supported by the twin domains observed with a polarized microscope [Fig. \ref{fig:BMX_lattice}(b)].
Note here that a similar distorted square net is formed in $A$MnSb$_2$ ($A\!\ne$Ba) as well\cite{Liu2017NatMater,Ramankutty2018SciPost,You2018CAP,He2017PRB}.
However, the type of layer stacking is totally different; BaMn$X_2$ exhibits the coincident arrangement of Ba atoms above and below the $X^-$ layer, whereas $A$MnSb$_2$ ($A\!\ne$Ba) exhibits the staggered arrangement.
The former has a new crystal structure (space group $Imm2$) among the $A$Mn$X_2$ family, which is polar along the in-plane direction [thick arrow along the $c$ axis in Fig. \ref{fig:BMX_lattice}(a)], whereas the latter has a known nonpolar structure (space group $Pnma$).
%
\par
%
To verify the polar lattice structure directly, we measured the optical second harmonic generation (SHG).
SHG is the frequency doubling of the probing light wave and, to leading order, occurs only in non-centrosymmetric systems [Fig. \ref{fig:BMX_lattice}(c)].
In Fig. \ref{fig:BMX_lattice}(d), we show the rotational anisotropy of the SHG intensity from a nearly single-domain region on a cleaved surface.
The rotational anisotropy was obtained by projecting the component of the SHG light oriented parallel to the polarization $\phi$ of the incident fundamental light ($\phi$).
By rotating $\phi$ over $360^{\circ}$, we observed a clear twofold anisotropy with peaks at $\phi\sim90^{\circ}$ and $270^{\circ}$ [Fig. \ref{fig:BMX_lattice}(d)], which is consistent with the fitting based on the $mm2$ symmetry with in-plane lattice polarization along the $c$-axis [solid curve in Fig. \ref{fig:BMX_lattice}(d)]\cite{Kondo2021CommunMater}.
%
\par
%
Interestingly, the magnitude of polar lattice distortion in BaMn$X_2$ largely depends on the $X$ species.
From the results of single-crystal X-ray structural analysis, the orthorhombicity $(c-a)/a$ of BaMnBi$_2$ is estimated to be $\sim\!0.15\%$\cite{Kondo2021CommunMater}, which is roughly $1/10$ that of BaMnSb$_2$ ($\sim\!1.3\%$)\cite{Sakai2020PRB}.
Such a difference in orthorhombicity results in a marked impact on the Dirac bands and Fermi surfaces, as detailed in Sect. \ref{sec:BMX_SV}.
Below, to explain the fundamental electronic and transport properties of BaMn$X_2$, we first focus on BaMnSb$_2$ with the larger orthorhombicity.
%
%
\subsection{Valley-contrasting spin polarization in Dirac band}
\begin{figure}[tb]
\begin{center}
\includegraphics[width=.8\linewidth]{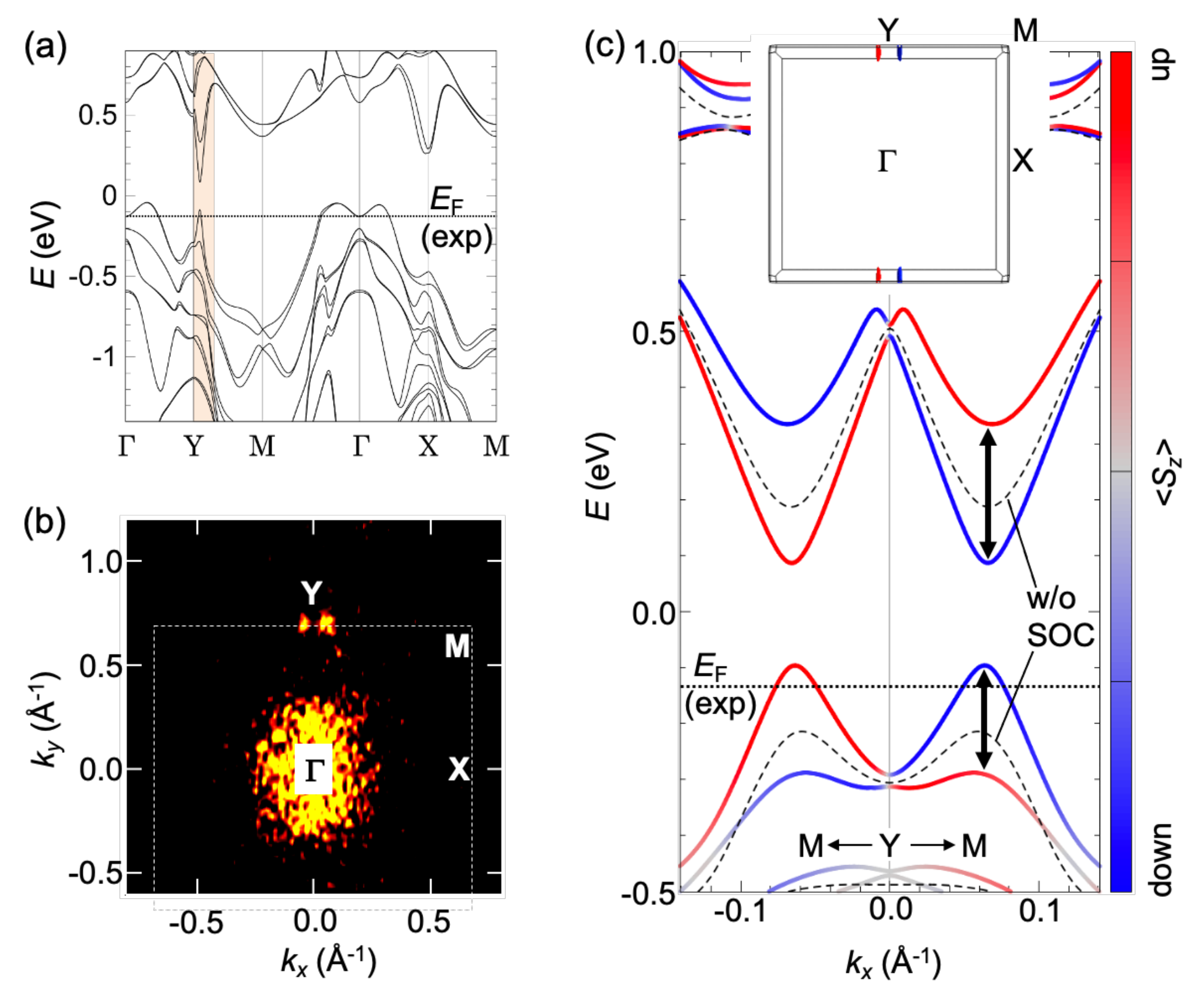}
\end{center}
\caption{(Color online)
(a) Calculated band structure for BaMnSb$_2$ with the SOC.
The Fermi energy $E_{\rm F}$ ($=-0.13$ eV) is estimated from the experimental SdH oscillation [see caption of (c)].
(b) Fermi surfaces for BaMnSb$_2$ deduced from the ARPES intensity map at $\sim$ 60 K (integrated over 100 meV at the Fermi level).
The $k$-component along the crystallographic $a$($c$)-axis corresponds to $k_x$($k_y$).
(c) Calculated Dirac bands near the Y point along the M-Y-M line, where the $\langle S_z \rangle$ value of each band is represented by red (up) and blue (down) colors.
The dashed gray curves denote the bands without the SOC.
The vertical arrows indicate the spin splitting induced by the SOC.
The inset shows the spin-resolved Fermi surface, where $E_{\rm F}$ ($=-0.13$ eV) was determined so that the valley size matches the result of the SdH oscillation.
Note here that the large Fermi surface located at the $\Gamma$ point [panel (a)] is neglected, since it likely stems from the underestimation of the band gap\cite{note_artifact}.
Reproduced with permission from Refs. \onlinecite{Sakai2021Butsuri} (\copyright 2021 Physical Society of Japan) and \onlinecite{Sakai2020PRB} (\copyright 2020 American Physical Society) and edited by the author.
}
\label{fig:BMX_band}
\end{figure}

%
Figure \ref{fig:BMX_band}(a) shows the band structure of BaMnSb$_2$ calculated on the basis of the experimental crystal structure.
For tetragonal $A$Mn$X_2$, (e.g., SrMnBi$_2$~\cite{Park2011PRL,Lee2013PRB} and EuMnBi$_2$~\cite{Borisenko2019NatCom}), the Dirac band along the $\Gamma$-M line crosses the Fermi energy, forming four equivalent valleys.
However, for orthorhombic BaMnSb$_2$, the zigzag-type distortion leads to a considerable gap ($\sim 1$ eV) at the Dirac points on the $\Gamma$-M lines.
Another Dirac band can form at the X(Y) point, as shown in Fig. 1(b).
For (Sr,Eu)MnBi$_2$, the SOC-induced energy gap is so large that no Fermi pockets are created\cite{Lee2013PRB,Borisenko2019NatCom}.
For BaMnSb$_2$, on the other hand, although a large gap is formed at the X point, the energy gap around the Y point remains relatively small ($\sim$200 meV), leading to a highly dispersive gapped Dirac band [shaded area in Fig. \ref{fig:BMX_band}(a)].
Its valence band crosses the Fermi level owing to a few hole carriers doped in real crystals\cite{note_artifact}.
Consequently, two small hole pockets form near the Y point along each Y-M line, as was observed in the ARPES experiments [Fig. \ref{fig:BMX_band}(b)].
Note here that the broad intensity around the $\Gamma$ point arises from the less dispersive parabolic valence band located just below the Fermi level.
Although finite intensity appears in the ARPES mapping as a result of energy integration over 100 meV, this band is unlikely to form a Fermi surface in reality, which is also supported by transport features (see Sect. \ref{sec:BMX_QHE}).
%
\par
%
%
We now discuss the role of the SOC as a source of the peculiar spin polarization of Dirac bands for BaMnSb$_2$.
Figure \ref{fig:BMX_band}(c) displays the spin-resolved band structure along the M-Y-M line near the Y point, where the $\langle S_z\rangle$ value is denoted by red (spin-up) and blue (spin-down) colors.
Although the Dirac band is spin-degenerate without the SOC (dashed curve), it exhibits $k_x$-contrasting large spin splitting when the SOC sets in.
This corresponds to the out-of-plane Zeeman-type splitting due to the polar structure within the plane\cite{Yuan2013NatPhys} and is qualitatively explained by the following SOC Hamiltonian: 
$H_{\rm SO}\propto\bm{\sigma}\cdot(\bm{k}\times\bm{E})\propto \sigma_zk_xE_y$ (for $k_z\! =\! 0$), where $\bm{\sigma}$ is the spin Pauli matrix and $\bm{E}\! =\! (0, E_y,0)$ is the built-in electric field due to the polar structure.
The energy splitting is so large [$\sim$200 meV denoted by the vertical arrows in Fig. \ref{fig:BMX_band}(c)] around the Dirac point that only the higher-energy valence band hosting a steep linear dispersion crosses the Fermi energy.
The resultant two small elliptic cylindrical Fermi surfaces along the M-Y-M line nicely reproduce the ARPES results, where each Fermi surface (valley) exhibits full $\langle S_z\rangle$ polarization opposite to one another [inset of Fig. \ref{fig:BMX_band}(c)].
%
\par
%
Such a spin-valley coupled Dirac band reminds us of monolayer transition metal dichalcogenide thin films such as MoS$_2$, which has a graphene-like structure with broken inversion symmetry.
In monolayer MoS$_2$, the electronic states forming the valleys at the K and K' points are described by a massive Dirac fermion exhibiting valley-contrasting spin polarization perpendicular to the 2D plane\cite{MoS2_valleytro1, MoS2_band_calc, MoS2_QO, MoS2_Xiao_PRL}.
Consequently, many distinctive optical and transport properties were theoretically predicted and experimentally observed, including valley-dependent circular dichroic photoluminescence\cite{MoS2_valleytro2, MoS2_monolayer_nanotech,MoS2_monoleyer_natcommun} and nonreciprocal charge transport\cite{MoS2_nonreci}.
However, the spin-valley coupling in MoS$_2$ disappears in the most stable bulk form (so-called 2H polytype), because the inversion symmetry is restored by the compensation between the layers\cite{JAWilson_TMDC}.
For BaMn$X_2$, on the other hand, each layer has the same polarity and thus the spin-valley coupling is realized in a bulk form, which is highlighted in the relativistic QHE in three dimensions shown below.
%
\subsection{Bulk quantum Hall effect}
\label{sec:BMX_QHE}
\begin{figure}[tb]
\begin{center}
\includegraphics[width=\linewidth]{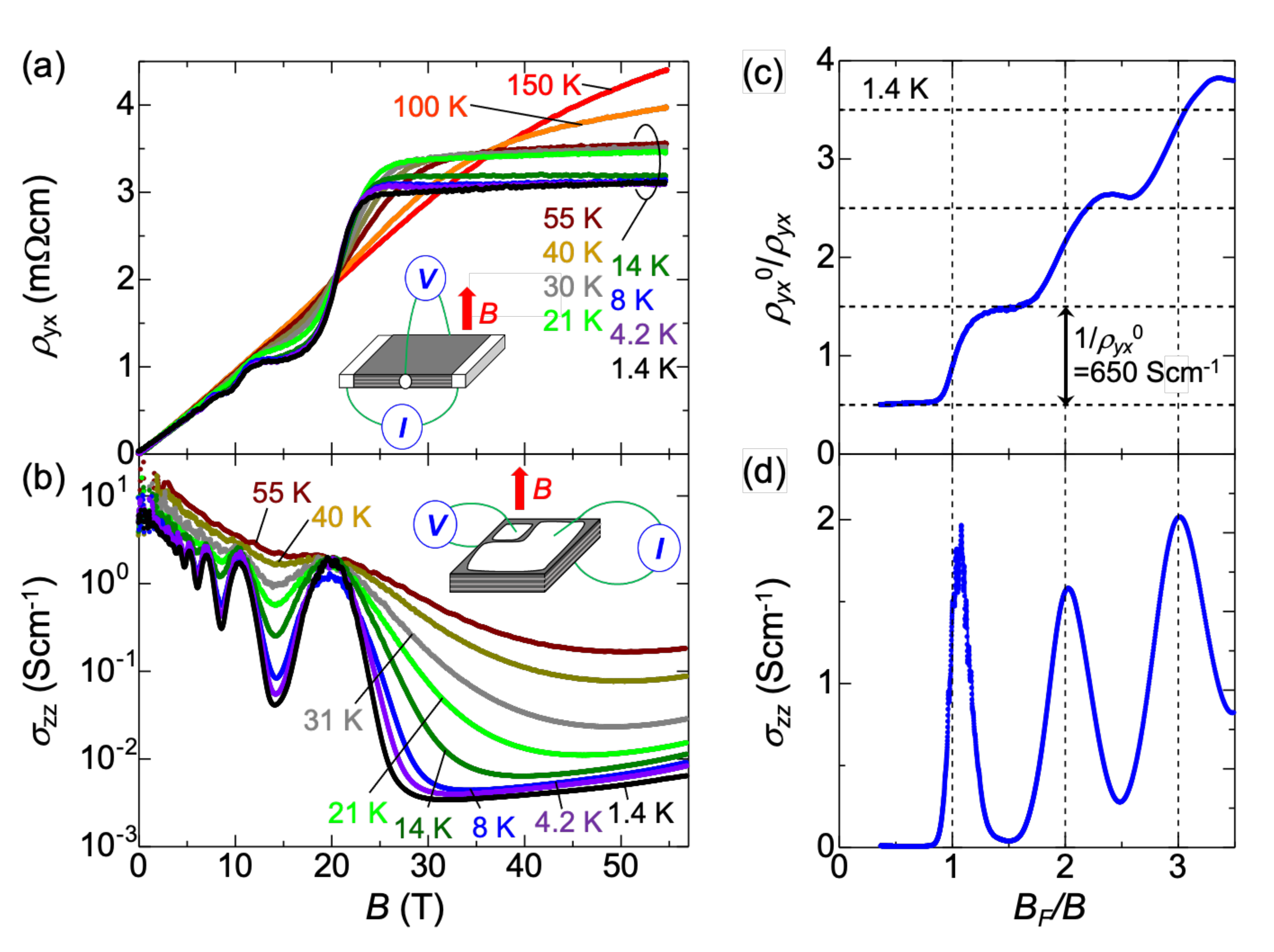}
\end{center}
\caption{(Color online)
Field ($B$) dependences of (a) Hall resistivity $\rho_{yx}$ and (b) interlayer conductivity $\sigma_{zz}$ ($=\!1/\rho_{zz}$) for BaMnSb$_2$ at various temperatures for $B||b$ up to $\sim$55 T.
The measurement setup is schematically shown in each panel.
(c) Normalized inverse Hall resistivity ($\rho_{yx}^0/\rho_{yx}$) and (d) $\sigma_{zz}$ versus $B_F/B$ at 1.4 K for $B||b$, where $B_F$(=21.3 T) is the SdH frequency.
$1/\rho_{yx}^0$(=650 Scm$^{-1}$) is obtained from the step size of $1/\rho_{yx}$ plateaus between $B_F/B\! =\! 0.5$ and 1.5, where $\sigma_{zz}$ is almost vanishing.
Reproduced with permission from Ref. \onlinecite{Sakai2020PRB} (\copyright 2020 American Physical Society) and edited by the author.
}
\label{fig:BMX_QHE}
\end{figure}

%
Figure \ref{fig:BMX_QHE}(a) shows the field ($B$) dependence of $\rho_{yx}$ for BaMnSb$_2$ at various temperatures for the field perpendicular to the plane ($B||b$).
$\rho_{yx}$ is almost linear with respect to $B$ up to 55 T above 150 K, consistent with the single hole carrier.
At 100 K, however, $\rho_{yx}$ weakly bends downward above 40 T, which evolves into a distinct plateau structure below 55 K.
In addition to this main $\rho_{yx}$ plateau (above 25 T), other $\rho_{yx}$ plateaus at lower fields become clear as temperature further decreases.
Correspondingly, we observed giant SdH oscillations in $\sigma_{zz}$ at low temperatures [Fig. \ref{fig:BMX_QHE}(b)]; the ratio of the peak $\sigma_{zz}$ value (at $\sim$20 T) to the bottom one (at $\sim$30 T) reaches $\sim$600 at 1.4 K.
Such vanishing $\sigma_{zz}$ accompanied by the $\rho_{yx}$ plateau is a clear hallmark of the QHE.
Although the precise measurement of $\rho_{xx}$ in a bulk single crystal was impossible because of the mixture of interlayer and in-plane voltage drops, a similar QHE has recently been reported by measuring $\rho_{xx}$ in a microfabricated crystal\cite{Liu2021NatCom}.
%
\par
%
%
In Figs. \ref{fig:BMX_QHE}(c) and \ref{fig:BMX_QHE}(d), $1/\rho_{yx}$ and $\sigma_{zz}$ at 1.4 K are plotted as a function of $B_F/B$, respectively, where $B_F$ is the frequency of the SdH oscillation ($=\!21.3(2)$ T).
Clear $1/\rho_{yx}$ plateaus and deep $\sigma_{zz}$ minima manifest themselves near half-integer values of $B_F/B$.
In particular, $\sigma_{zz}$ is almost zero for $B_F/B\!=\! 0.5$ and $1.5$, reflecting the well-defined quantum Hall states at high fields.
Half-integers of $B_F/B$ indicate the nontrivial $\pi$ Berry phase in BaMnSb$_2$, as is the case for EuMnBi$_2$ [Fig. \ref{fig:EMB_Rzz}(c)].
Furthermore, the quantized values of $1/\rho_{yx}$ provide us with important information.
Assuming that each Sb layer stacked along the $b$-axis contributes to conduction in parallel, $\rho_{yx}$ is expressed by the Hall resistance of each layer ($R_{yx}$) as follows: $\rho_{yx}\! =\! R_{yx}/Z^\ast$, where $Z^\ast\! =\! 1/(b/2)\! =\! 8.23\times 10^6$ cm$^{-1}$ is the number of Sb layers per unit thickness with $b$ being the $b$-axis length\cite{Huang2016PNAS, Sakai2020PRB}.
$R_{yx}$ should be quantized as $1/R_{yx}\! =\!\pm g_sg_v\left(N\! +\! 1/2\! -\!\gamma\right)e^2/h$~\cite{Zheng2002PRB,Gusynin2005PRL}, where $e$ is the electronic charge, $h$ is Planck's constant, and $g_s$ ($g_v$) is the spin (valley) degeneracy factor.
From these relations, $1/\rho_{yx}$ plateaus are given by $1/\rho_{yx}\!=\! Z^\ast g_sg_v(N\! +\! 1/2\! -\!\gamma)e^2/h$, where $Z^\ast g_sg_ve^2/h\!\equiv\! 1/\rho_{yx}^0$ corresponds to the step size between adjacent plateaus.
Figure \ref{fig:BMX_QHE}(c) shows $1/\rho_{yx}$ scaled by $1/\rho_{yx}^0$, where the plateaus are nicely quantized to half-integer values, i.e., $N\! +\! 1/2\! -\!\gamma\!=\! 0.5$, 1.5, 2.5,....
This clearly supports that each Sb layer exhibits the half-integer QHE ($\phi_B\! =\! \pi$), characteristic of the Dirac fermion.
Furthermore, from the experimental value of $1/\rho_{yx}^0\! =\! 650\!\pm\! 60$ Scm$^{-1}$, the total spin-valley degeneracy factor $g_sg_v$ is estimated to be $2.0\!\pm\! 0.2$.
This value is consistent with the two spin-polarized Dirac valleys (i.e., $g_s\!=\!1$ and $g_v\!=\!2$) shown in Figs. \ref{fig:BMX_band}(b) and \ref{fig:BMX_band}(c).
%
\par
%
Similar analyses on the bulk QHE in EuMnBi$_2$ can be performed using the high-field resistivity data [Figs. \ref{fig:EMB_highB}(c)--(e)], although the plateaus of $\rho_{yx}$ or the minima of $\sigma_{zz}$ are not as prominent as those in BaMnSb$_2$.
From Fig. \ref{fig:EMB_highB}(e), $1/\rho_{yx}^0$ is estimated to be 17,000-19,000 Scm$^{-1}$, resulting in $g_sg_v\!=\!5$-6~\cite{Masuda2016SA}.
This $g_sg_v$ value is roughly consistent with the spin-degenerate ($g_s$=2) four Dirac valleys located on each $\Gamma$-M line ($g_v$=4) for EuMnBi$_2$\cite{Borisenko2019NatCom, Kondo2020JPSJ}.
Thus, $A$Mn$X_2$ exhibits QHE in a bulk form at high fields.
Its detailed analyses uncover not only the nontrivial Berry phase but also the spin-valley state of the Dirac fermion, which clearly reflects the coupling with the magnetic order and lattice polarization.
%
\par
\begin{figure}[tb]
\begin{center}
\includegraphics[width=.6\linewidth]{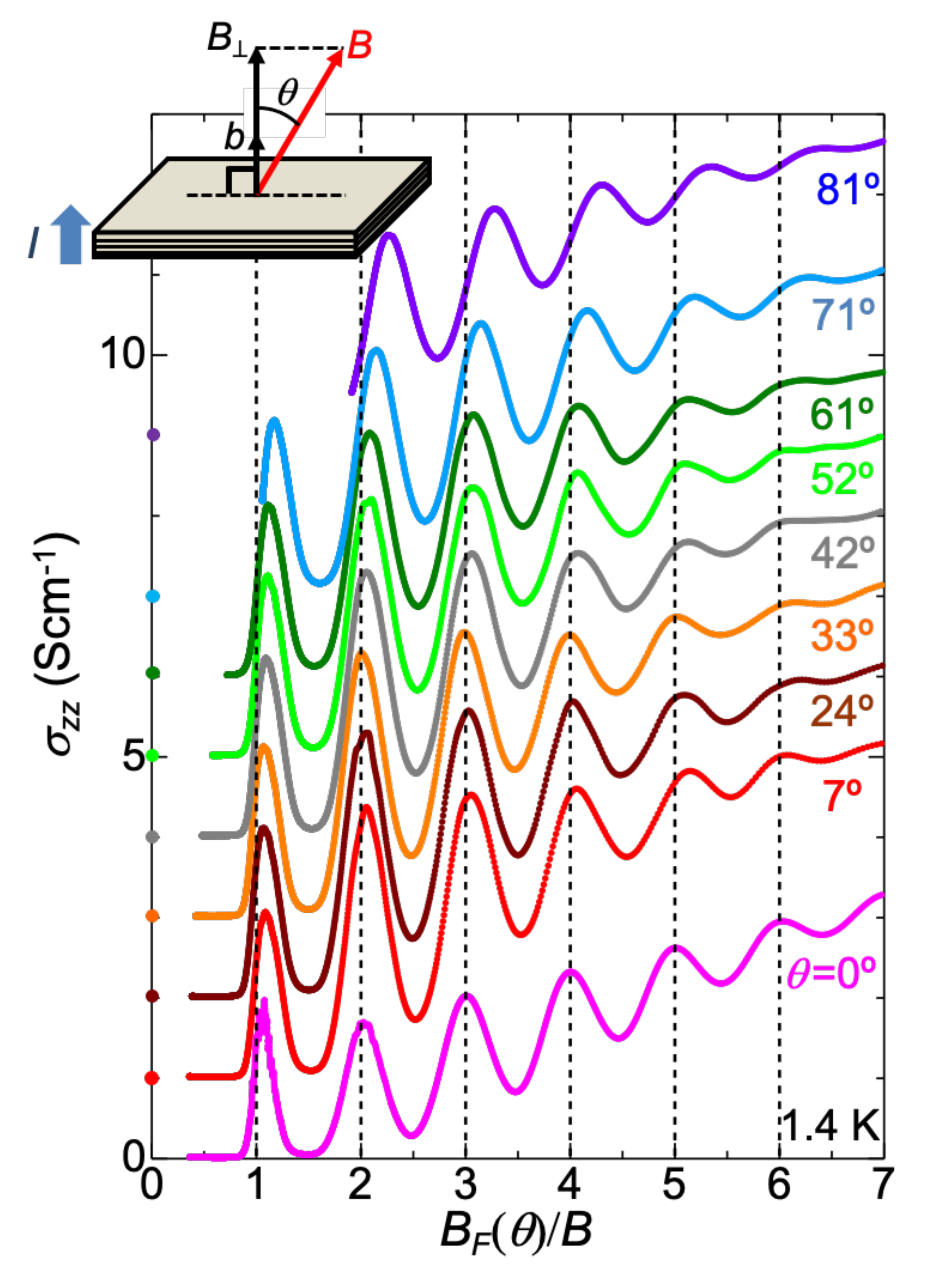}
\end{center}
\caption{(Color online)
Angular ($\theta$) dependence of $\sigma_{zz}$ versus $B_F(\theta)/B$ at 1.4 K, where $B_F(\theta)$ is the SdH frequency at each $\theta$ [$B_F(\theta\!=\! 0^\circ)\!=\! 21.3$ T].
$\theta$ is the angle between the field and the $b$ axis (inset). Each curve at $\theta\!\ge 7^\circ$ is shifted vertically (by 1 or 2 Scm$^{-1}$) for clarity.
Each origin is denoted by a closed circle on the left axis.
Inset: geometry of the interlayer resistivity measurement in tilted fields.
Reproduced with permission from Ref. \onlinecite{Sakai2020PRB} (\copyright 2020 American Physical Society) and edited by the author.
}
\label{fig:BMX_angle}
\end{figure}

%
To further demonstrate the spin polarization of the Dirac fermion, we measured the variation of SdH oscillation upon field angle ($\theta$) for BaMnSb$_2$.
As explained in Sect. \ref{sec:EMB_Rzz}, when $E_{\rm Z}$ is increased to $E_{\rm c}/2$ by increasing $\theta$ in a 2D system, the amplitude and phase of SdH oscillation should change significantly as a result of the overlap of the neighboring Landau levels with opposite spins\cite{Si100_spinsucceptibility_Fang1968, Masuda2018PRB}.
Figure \ref{fig:BMX_angle} shows the $\theta$ dependence of $\sigma_{zz}$ versus $B_F(\theta)/B$ at the lowest temperature, where $B_F(\theta)$ is the SdH frequency determined at each $\theta$.
$B_F(\theta)$ increases with $\theta$ by following the $1/\cos\theta$ scaling up to $\theta$=81$^\circ$, reflecting the strong 2D nature of the Fermi surface (data not shown here)\cite{Sakai2020PRB}.
Notably, the overall profiles of SdH oscillation plotted against $B_F(\theta)/B$ are almost independent of $\theta$, except for a slight phase shift above 71$^\circ$\cite{Tang2019Nature}.
We thus observed no spin splitting, as is consistent with fully spin-polarized Dirac valleys.
One should note that, since two Dirac valleys are oppositely spin-polarized, the remaining valley degeneracy should be lifted by $E_{\rm Z}$ when the field is applied along the $S_z$ axis ($\theta$=0$^\circ$).
However, no signature of splitting of the Landau levels is discernible at $\theta$=0$^\circ$ (Fig. \ref{fig:BMX_angle}), which may indicate that the SOC-induced energy splitting is much larger than $E_{\rm Z}$ in BaMnSb$_2$.
If this is the case, the spins for the Dirac valleys are almost locked along the $S_z$ axis even when the field is tilted, leading to $E_{\rm Z}\!\propto\! B_\perp$.
Therefore, $E_{\rm Z}/E_{\rm c}$ is nearly independent of $\theta$ and reproduces the experimental results shown here.
%
%
\subsection{Tunable spin-valley coupling}
\label{sec:BMX_SV}
%
The valley configuration plays an important role in the peculiar optoelectronic and transport phenomena in the spin-valley coupled systems\cite{MoS2_valleytro1, MoS2_band_calc, MoS2_QO, MoS2_Xiao_PRL, MoS2_valleytro2, MoS2_monolayer_nanotech,MoS2_monoleyer_natcommun,MoS2_nonreci}.
However, the valley position for monolayer MoS$_2$ is fixed at the corners of the hexagonal Brillouin zone (the K and K' points)\cite{MoS2_band_calc, MoS2_Xiao_PRL}, reflecting the graphene-like structure.
Contrary to this, the spin-valley state for BaMn$X_2$ is widely tunable by the chemical substitution of the $X$ site ($X$=Sb, Bi), as detailed below.
%
\par
\begin{figure}[tb]
\begin{center}
\includegraphics[width=.6\linewidth]{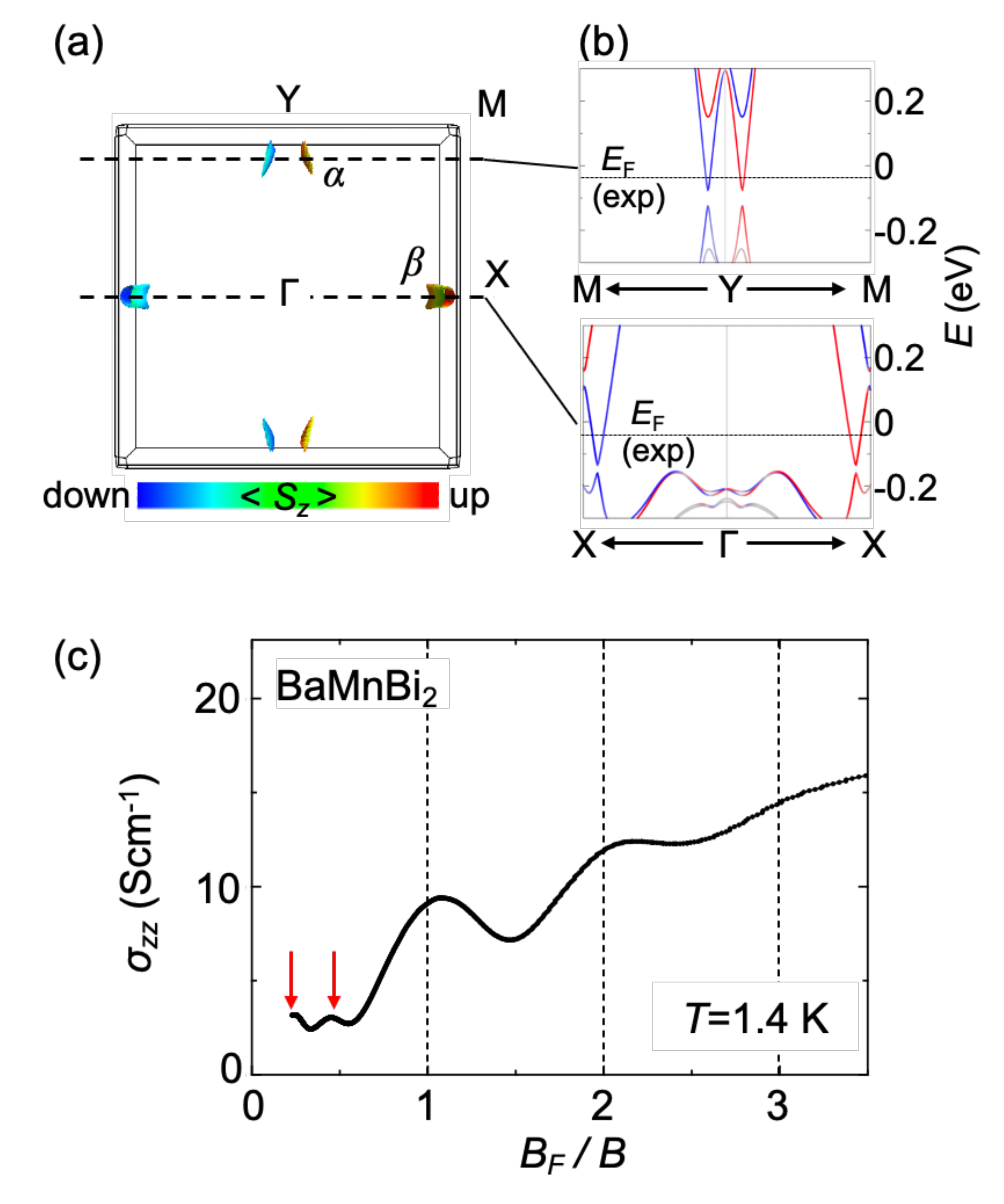}
\end{center}
\caption{(Color online)
(a) Spin-resolved Fermi surfaces for BaMnBi$_2$ ($E_F\!=\!-33$ meV).
The color of the Fermi surface represents the spin polarization of $\left<S_z\right>$.
In calculation, hole valleys appear on the $\Gamma$-M line, but they are neglected here since there is no experimental signature for them\cite{Kondo2021CommunMater}.
(b) Band dispersion cut along the dashed lines shown in panel (a).
The red and blue colors represent spin up and down, respectively.
The horizontal dotted line indicates the experimental $E_F$ value (=$-33$ meV), which is determined so that the size of the $\alpha$ valley coincides with the $B_F$ value (=13 T) of the main SdH oscillation in panel (c). 
(c) $\sigma_{zz}$ versus normalized filling factor $B_F/B$ for BaMnBi$_2$ at 1.4 K.
The vertical arrows indicate an additional oscillation structure observed at high fields.
Reproduced with permission from Ref. \onlinecite{Kondo2021CommunMater} (\copyright 2021 The authors) and edited by the author.
}
\label{fig:BMX_SV}
\end{figure}

%
As a result of the synchrotron X-ray diffraction, we found that BaMnBi$_2$ has a polar orthorhombic structure similar to that of BaMnSb$_2$, but its orthorhombicity is as small as one-tenth of that for BaMnSb$_2$\cite{Kondo2021CommunMater}.
The corresponding change in spin-valley state is revealed by calculating the band structure of BaMnBi$_2$ based on the experimentally determined crystal structure.
Figure \ref{fig:BMX_SV}(a) shows the calculated spin-resolved Fermi surfaces, where $E_F$ is set to 33 meV so as to reproduce the SdH oscillation [Fig. \ref{fig:BMX_SV}(c)]. 
Around the X and Y points, 2D cylindrical Fermi surfaces associated with the Dirac bands are formed, two of which ($\beta$ valleys) are near the X point and four of which ($\alpha$ valleys) are near the Y point.
To show the spin polarization of each valley, we show in Fig \ref{fig:BMX_SV}(b) the spin-resolved band dispersion along the $k_y=0.8\pi/c$ and $k_y=0$ lines that are located at the centers of the $\alpha$ and $\beta$ valleys, respectively.
In both valleys, the magnitude of spin splitting is so large (200--300 meV) owing to the strong SOC in Bi that one of the spin-split Dirac bands crosses $E_F$, resulting in the full polarization of $\left<S_z\right>$. 
The polarity of $\left<S_z\right>$ switches sign with respect to the $\Gamma$-Y line, similarly to BaMnSb$_2$.
Reflecting the smaller lattice distortion for BaMnBi$_2$, the energy gap ($\sim50$ meV) at the Dirac points is smaller than that for BaMnSb$_2$ ($\sim200$ meV), forming nearly massless Dirac bands.
Thus, the Dirac valleys ($\alpha$ and $\beta$) for BaMnBi$_2$ [Fig. \ref{fig:BMX_SV}(a)] are totally different from those for BaMnSb$_2$ [inset of Fig. \ref{fig:BMX_band}(c)], suggesting that the spin-valley coupling in this series of materials is highly sensitive to the lattice distortion and SOC.
%
%
\par
%
The multiple-valley state for BaMnBi$_2$ affects the Landau level structure at high fields.
In Fig. \ref{fig:BMX_SV}(c), we plot $\sigma_{zz}$ as a function of $B_F/B$ ($B_F\!=\!13$ T) for BaMnBi$_2$.
$\sigma_{zz}$ shows clear dips at half-integers of $B_F/B$, consistent with the nontrivial $\pi$ Berry phase.
Importantly, at $B_F/B\!<\!1$ corresponding to the zeroth Landau level, an additional oscillation structure [vertical arrows in Fig. \ref{fig:BMX_SV}(c)] appears, which cannot be explained by the single SdH frequency, nor is detected in BaMnSb$_2$ [Fig. \ref{fig:BMX_QHE}(d)].
Given that BaMnBi$_2$ has two inequivalent Dirac valleys ($\alpha$ and $\beta$ in Fig.\,2b), the additional structure may stem from the SdH oscillation of another valley superimposed on the main SdH oscillation.
In fact, the Fermi surface sizes estimated by assuming the two SdH frequencies are semiquantitatively reproduced by first-principles calculation for $E_F\!=\! -33$ meV [Fig. \ref{fig:BMX_SV}(b)]\cite{Kondo2021CommunMater}.
It is thus likely that the additional oscillation at $B_F/B<1$ is relevant to the multiple Dirac valleys in BaMnBi$_2$.
However, to clarify the exact origin, further studies on the Fermi surface, such as ARPES, are necessary as a future work.
%
\section{Summary and Outlook}
\label{sec:summary}
%
In this review, we have given an overview of the Dirac fermion coupled with magnetic order and lattice polarization in the layered material $A$Mn$X_2$, where the systematic exploration of Dirac metals has been demonstrated on the basis of the block-layer concept.
Taking advantage of the high mobility of the Dirac fermion layer spatially separated from the block layer, we have studied the quantum oscillation and bulk quantum Hall effect by high-field transport measurements using a pulsed magnet.
In combination with first-principles calculations, we have clarified the large spin splitting and strong spin-valley locking of the quasi 2D Dirac band for $magnetic$ EuMnBi$_2$ and $polar$ BaMn$X_2$, respectively.
%
\par
%
Importantly, since such a spin-valley-polarized Dirac fermion is realized in a bulk form, its state is tunable by various methods.
In EuMnBi$_2$, we have found that the interlayer conductivity, as well as the magnitude of spin splitting, markedly depends on the magnetic order of Eu layers, which is controllable via an external magnetic field.
We have also demonstrated the control of the spin-valley state for BaMn$X_2$ via the chemical substitution of the $X$ site as a result of the modulation of the lattice distortion and spin-orbit coupling.
Furthermore, the partial substitution of the divalent $A$ site with a trivalent one in the block layer leads to electron doping in the Dirac fermion layer, enabling Fermi level tuning across the charge neutral point (not reviewed here in detail)\cite{Tsuruda2021AdvFunctMater}. 
In addition to magnetic field and chemical substitution, it has been recently predicted that mechanical strain could also significantly affect the band structure and/or topology owing to the high sensitivity to lattice distortion\cite{Yu2020NatCom}.
The widely tunable Dirac fermion characteristic of $A$Mn$X_2$ would be essential for optimizing their exotic bulk properties and helpful for future device application.
%
\begin{acknowledgments}
The author thanks all the collaborators of this research for their various contributions.
This work was partly supported by JST PRESTO (Grant No. JPMJPR16R2), JSPS KAKENHI (Grant Nos. 16H06015, 19H01851, 19K21851, and 22H00109), and the Asahi Glass Foundation.
\end{acknowledgments}
%
%
%


\begin{thebibliography}{100}
\bibitem{note_cell_doubling} Two sets of bands arise from the unit cell doubling along the $c$-axis to represent the AFM order of Eu moments, which is also adopted to the forced FM state in common.
\end{thebibliography}
\end{document}